\documentclass[a4paper,twoside]{article}
\pdfoutput=1
\usepackage{hyperref}
\usepackage{amsmath}
\usepackage{bm}
\usepackage{url}
\usepackage{booktabs}
\usepackage{epsfig}
\usepackage{subcaption}
\usepackage{calc}
\usepackage{amssymb}
\usepackage{amstext}
\usepackage{amsmath}
\usepackage{amsthm}
\usepackage{multicol}
\usepackage{pslatex}
\usepackage{apalike}
\usepackage{algorithm2e}
\usepackage[bottom]{footmisc}
\usepackage{SCITEPRESS}     

\begin{document}

\title{Addressing Label Leakage in Knowledge Tracing Models}

\author{\authorname{Yahya Badran\sup{1,2}\orcidAuthor{0009-0006-9098-5799}, Christine Preisach\sup{1,2}\orcidAuthor{0009-0009-1385-0585}}
\affiliation{\sup{1}University of Applied Sciences, Moltekstr. 30, 76133 Karlsruhe, Germany }
\affiliation{\sup{2}Karlsruhe University of Education, Bismarckstr 10,76133 Karlsruhe, Germany}
\email{\{yahya.badran, christine.preisach\}@h-ka.de}
}


\abstract{Knowledge Tracing (KT) is concerned with predicting students' future performance on learning items in intelligent tutoring systems. Learning items are tagged with skill labels called knowledge concepts (KCs). Many KT models expand the sequence of item-student interactions into KC-student interactions by replacing learning items with their constituting KCs. This approach addresses the issue of sparse item-student interactions and minimises the number of model parameters. However, we identified a label leakage problem with this approach. The model's ability to learn correlations between KCs belonging to the same item can result in the leakage of ground truth labels, which leads to decreased performance, particularly on datasets with a high number of KCs per item.

In this paper, we present methods to prevent label leakage in knowledge tracing (KT) models. Our model variants that utilize these methods consistently outperform their original counterparts. This further underscores the impact of label leakage on model performance. Additionally, these methods enhance the overall performance of KT models, with one model variant surpassing all tested baselines on different benchmarks. Notably, our methods are versatile and can be applied to a wide range of KT models.
}

\onecolumn \maketitle \normalsize \setcounter{footnote}{0} \vfill

\section{\uppercase{Introduction}}

Knowledge tracing (KT) models are essential for personalization and recommendation in intelligent tutoring systems (ITSs). Furthermore, some KT models can provide mastery estimation of the skills or concepts covered in the coursework. These concepts are typically listed by the ITS and referred to as Knowledge Concepts (KCs). Each question in the coursework can be assigned a set of KCs that are required to pass it. For example, a simple question such as "$1 + 5 - 3 =  ?$" might be tagged with two KCs: "summation" and "subtraction". The mastery estimation of KCs can form a state representation of the student at a point in time which can be the basis for a recommendation algorithm.
Many KT models use KCs to address the issue of data sparsity, which is due to the large number of questions available in ITS and the limited student-question interactions\cite{pykt,akt}. To achieve this, each question can be unfolded into its constituting KCs, creating a new KC-student interaction sequence instead of question-student interactions. Since the number of KCs is relatively much smaller than the number of questions, this approach helps mitigate the sparsity problem and also minimizes the number of parameters required in the model. 

When using the KC-student sequence, KCs of the same question form a subsequence. In production settings, the subsequence labels are either fully known (if the student responded to the question) or fully unknown (if the student has not interacted yet with the question). To accurately evaluate models trained on the KC-student sequence, it is crucial to apply techniques that replicate production settings. Failing to do so can result in ground-truth label leakage during evaluation, leading to misleading results\cite{pykt}. Unfortunately, this issue is often overlooked in the literature, causing reported performance metrics to be artificially inflated compared to the model's true performance\cite{pykt}.


Moreover, we noticed that models trained using this method can learn to leak ground-truth labels between KCs of the same question instead of inferring predictions based on the preceding questions in the sequence. As a result, models trained using this approach experience a decline in performance. The underlying reason is that deep learning models can learn to infer if KCs belong to the same question or not and, thus, learn to leak ground-truth labels between KCs of the same question. This problem is more pronounced with datasets containing a high average number of KCs per question because such datasets are more likely to contain correlated KCs (KCs that are more likely to occur together in the same question). 

It is important to note that classical models, such as \textit{Bayesian Knowledge Tracing} (BKT)\cite{bkt}, employ a strong independence assumption between knowledge components (KCs)\cite{BKTindep,abdelrahman2023knowledge}. These models do not learn dependencies between KCs and therefore do not suffer from this label leakage problem. While these classical models are more interpretable, their performance often does not match that of deep learning methods \cite{Howdeep,whendeep,DKTforget}. Accordingly, this paper concentrates exclusively on Deep Learning Knowledge Tracing (DLKT) models, which are capable of learning intricate dependencies between KCs.

To address the label leakage problem and show its effect, we eliminate any computational path that could leak ground truth labels during both training and evaluation using different methods. One method that we propose replaces ground-truth labels with a $\texttt{MASK}$ label whenever leakage might occur, inspired by masked language modeling \cite{BERT}. The advantage of this method is that it can be applied to various architectures. Our model variants that employ these methods significantly outperform their original counterparts, which highlights the impact of label leakage.

 Evaluating models on the KC-student interaction sequence can suffer yet from another problem. The length of the KC-student interaction sequence is longer than the question-student sequence, which is ignored in benchmark comparisons that do not enforce a fixed sequence length of questions across different KT models. Most benchmarks use datasets with a small average number of KCs per question, hence, the difference in length between the expanded KC sequence and the original question sequence is usually small for such datasets. However, once the dataset has larger average KCs per question, the model that uses the KC-student sequence is evaluated on sequences with fewer questions, which can result in unfair comparison.

This paper provides the following contributions:
\begin{itemize}
    \item We provide empirical evidence for ground-truth label leakage in commonly used KT models.
    \item We introduce a number of methods to prevent label leakage.
\item Our model variants that utilize the introduced methods exhibit competitive performance, winning different benchmarks.
    \item We used datasets with varying average number of KCs per question and use the same sequence lengths across different models for a fair comparison. 
    \item We publish our implementation as an open source tool: \url{https://github.com/badranx/KTbench}
\end{itemize}






. 



\section{\uppercase{Related Work}}
Originally knowledge tracing was mostly based on educational theories such as Item Response Theory (IRT) and mastery learning \cite{abdelrahman2023knowledge}. One example of such models is BKT which learns a Hidden Markov Model, where each skill corresponds to two states, one indicating mastery and the other not. These models do not represent any dependencies between different KCs and thus they do not suffer from the problem of label leakage discussed in this paper. However, this comes at the expense of being less expressive and typically resulting in lower performance overall.

Deep learning is capable of capturing more complex relations. The first DLKT model was Deep Knowledge Tracing (DKT)\cite{dkt} which uses recurrent neural networks (RNNs) to model the sequence of student interactions. Later, more DKT variants were introduced \cite{dkt_plus,DKTforget}. However, most of these models operate at the KC level on an expanded sequence and thus they can suffer from label leakage.  

\cite{pykt} discussed label leakage issues during evaluation. They proposed a method that mimic production settings to effectively evaluate KT models. However, they did not address the effect of label leakage on model performance. Moreover, their method is computationally expensive. Instead, we propose straightforward modifications to the original models that eliminate the need for specialised evaluation methods.  

Attention-based models, such as Transformers\cite{transformer}, offer a strong alternative to RNNs. Attention mechanisms require specialized masks to prevent unwanted data leakage within the sequence. For instance, preventing a model from accessing future information can be achieved using a simple triangular matrix. \cite{aaai2nd} presented a modified mask to prevent leakage between questions within each group. These groups are specific to their chosen dataset, in which students only receive feedback for the whole group instead of individual questions. This is done to properly model the student learning behavior, which is a special case for the used dataset. Similarly, we implement masks to prevent label leakage for attention based KT models that operate on the expanded KC sequence.

\section{\uppercase{Knowledge Tracing}}


Let $Q$ be the set of all questions. We represent the student interaction at time $t$ as a tuple $\left(q_t, r_t\right)$ where $q_t$ represents the question and $r_t$ represents the student response which is either $1$ if the answer is correct or $0$ otherwise. Let $\left(q_1, r_1\right), \ldots, \left(q_{t-1}, r_{t-1}\right)$ represent a chronological sequence of interactions by a single student up to time $t-1$. The main goal of a DLKT model is to predict $r_t$ for $q_t$,

\begin{equation}
r'_t = DLKT\left(q_t  ; (q_{t-1},  r_{t-1}), \ldots, (q_1, r_1)\right)
\end{equation}

where $r'_t$ is the model prediction.

Let $C$ be the set of all KCs. Since each question is tagged with a number of KCs, we represent the one-to-many mapping between each question and its KCs as $m: Q \rightarrow 2^C$ where $2^C$ is the set of all subsets of KCs.

\begin{figure}[h]
    \centering
\includegraphics[width=0.7\linewidth]{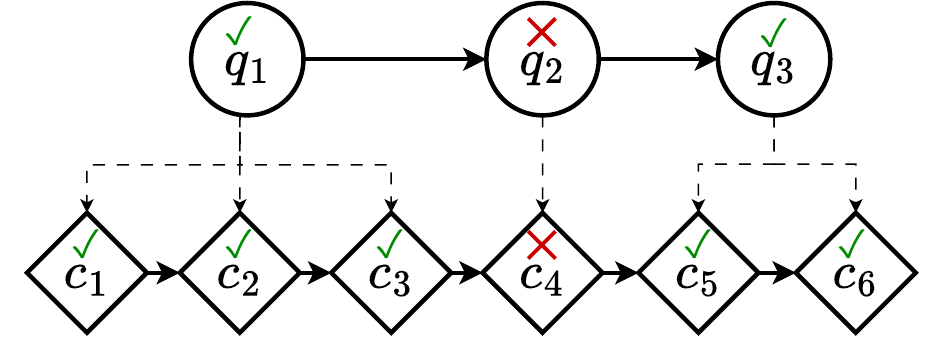}
\caption{Expanding a question-student interaction sequence into a KC-student interaction sequence. The green and red symbols are correct and incorrect respectively}\label{fig1}
\end{figure}

Many models expand the question-student interaction sequence into a KC-student interaction sequence. Each question in the sequence is expanded to its constituting KCs as illustrated in Figure~\ref{fig1}. For example, given an ordering over the KCs in $C$, let $q$ be a question with $n$ KCs then each interaction $\left(q, r\right)$ can be expanded into multiple interactions as follows: $\left(c_1, r\right) \ldots,(c_n, r)$ where $m(q) = \{c_1, c_2, \ldots, c_n\}$. This results in longer sequence lengths. Note that some models retain $q$ in the expanded sequence, while others drop it completely.

As the number of KCs is usually much smaller than the number of questions, this serves to mitigate the effect of sparse item-student interactions \cite{pykt}. Additionally, it can minimise the number of model parameters \cite{akt}. 
%
%
%

To illustrate this further, we review two important models that utilize this approach, DKT and AKT.

\subsection{Review of Deep Knowledge Tracing}
\label{dkt}
Deep Knowledge Tracing (DKT) was the first model to utilize deep learning for the knowledge tracing task \cite{dkt}. The model is mainly a recurrent neural network (RNN) with two variations: vanilla RNN and Long short-term memory (LSTM). In this work, we only consider the LSTM variant. The original implementation completely discards the questions and takes the expanded KC-student interaction sequence as input in order to predict a future response 

\begin{equation}
    r_t' = DKT\left(c_t ; (c_{t-1}, r_{t-1}), \ldots, (c_1, r_1)\right)
\end{equation}

Each KC-response pair, $(c,r)$, is mapped to a unique vector embedding $e_{(c, r)}$ in order to be processed by the recurrent model which in turns outputs a sequence of hidden states $\{h_t\}$ that are passed to a single-layer neural network as follows:

\begin{equation}
\mathbf{y}_t=\sigma\left(\mathbf{W} h_t+\mathbf{b}\right)
\end{equation}

 where $W$, $b$, and $\sigma$ are the weight, bias and sigmoid activation function respectively. The output $\mathbf{y_t}$ has a dimensionality equivalent to the cardinality of KCs such that each dimension represents the probability of a correct response for the corresponding KC. Thus, the model prediction would be $r'_t = \mathbf{y_t}[c_t]$, which is the value at dimension $c_t$.

The original implementation of DKT uses two methods to compute the vector $e_{(c, r)}$. One is a one-hot encoding of the tuple $(c, r)$, which means the resulting vector has a dimension of two times the number of KCs, $|C|$. Given the fact that this vector can have a very large dimension with higher number of KCs, they suggest to sample a random vector with fixed-dimension $d$ for each pair $(c, r)$, $e_{(c, r)} \sim \mathcal{N}(\mathbf{0}, \mathbf{I})$. In both methods, the vectors are fixed and not part of the learned parameters. 


Some recent implementations of DKT do not use these two methods \cite{pykt,edustudio}. Instead, each $(c, r)$ is mapped to a unique learned vector embedding of a fixed dimension $d$. Which is equivalent to passing a one-hot encoding to a linear layer with no bias which outputs $d$ features before passing it to the LSTM but at much lower computational cost. Thus, we similarly adopt this approach in our work.




%

%

%
%

\subsection{Review of Attentive Knowledge Tracing}
\label{akt} 
\textit{Attentive Knowledge Tracing }(AKT) \cite{akt} utilizes a self-attention mechanism to produce a contextualized representation of both, questions and student responses. Unlike scaled dot-product attention which depends on the order of the items in the sequence, AKT attention weights incorporate information about the relatedness between questions, which corresponds to increased attention weight for related questions in the sequence. It also incorporates student forgetting effects which corresponds to a decrease in attention weight with time (time is substituted by the order of the item in the sequence). They call this monotonic attention mechanism.

AKT has two self-attention based encoders. One is called the question encoder which is responsible for contextualized question representation. It takes embeddings computed from the input sequence without student response data $(q_{t-1}, c_{t-1}), \ldots, (q_1, c_1)$. The question encoder utilizes monotonic attention to output a contextualised representation, $x_t$, of the current question $q_t$

\begin{equation}
x_t = f_{qenc}(e_{(q_t, c_t)}, e_{(q_{t-1}, c_{t-1})}, \ldots, e_{(q_1, c_1)})
\end{equation}

Where $e_{(q_t, c_t)}$ is the embedding vector of $(q_t, c_t)$. The other encoder is called knowledge encoder which produces a contextualized student knowledge representation, $y_t$, as it takes embeddings computed from KCs and student response input data

\begin{equation}
y_t = f_{kenc}(e_{(r_{t-1}, c_{t-1}, q_{t_1})}, \ldots, e_{(r_1, c_1, q_1)})
\end{equation}

Where $e_{(r_t, c_t, q_t)}$ is the embedding vector of $(r_t, c_t, q_t)$. The outputs of both encoders are passed to a knowledge retriever, which utilizes a special monotonic attention mechanism to retrieve relevant past knowledge for the current question, 

\begin{equation}
h_t = f_{kr}(x_1, \ldots, x_t, y_1, \ldots, y_{t-1})
\end{equation}

\begin{figure}[h]
    \centering
\includegraphics[width=0.7\linewidth]{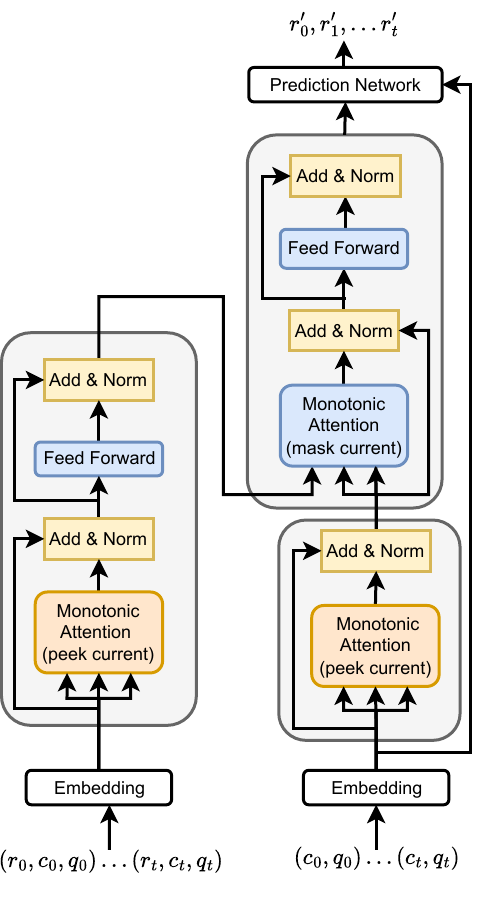}
\caption{Overview of the AKT model architecture. This is a simplified version, some blocks are repeated. Each attention block in the figure takes the sequence of inputs---value, query, and key---from left to right.}\label{fig:akt}
\end{figure}

Lastly, the output of the knowledge retriever, $h_t$, is passed to a feed-forward network to predict the question response on a specific KC. For an overview of the AKT architecture, see Figure \ref{fig:akt}.

AKT has two attention masks. The first is a lower triangular mask to prevent any connection between the output $x_t$ or $y_t$ and future items, $\{\left(r_{t+1}, c_{t+1}, q_{t+1}\right), \left(r_{t+2}, c_{t+2}, q_{t+2}\right), \ldots\}$, which is used by both encoders. The second is a strictly lower triangular mask (where the main diagonal contains zeros) to prevent any connection between $r'_t$ and both current and future items in the sequence which is used by the knowledge retriever. However, AKT still suffers from label leakage despite the provided attention masks. The reason is its use of the KC-student interaction sequence as input which we will explain in more details in section~\ref{problems}.



The embeddings are constructed using an approach inspired by the Rasch model which estimates the probability of a student answering a question correctly using two parameters: the difficulty of the question and the student's ability \cite{irt}. Using the Rasch model approach, it constructs two types of embeddings for both $(q_t, c_t)$  and $(r_t, c_t, q_t)$ as follows:

\begin{align}
 e_{(q_t,c_t)} & = e_{c_t} + \mu_{q_t} \cdot d_{c_t} \label{eq:akt_emb1}\\
 e_{(r_t, c_t, q_t)} & = e_{(c_t, r_t)} + \mu_{q_t} \cdot f_{(c_t, r_t)}  \label{eq:akt_emb2}
\end{align}

Both $d_{c_t}$ and $f_{(c_t, r_t)}$ are vector embeddings, called "variation vectors", while $\mu_{q_t}$ is a scalar representing question difficulty. $e_{c_t}$ is the embedding for each KC. Furthermore, the embedding of the concept-response pair is defined as 

\begin{equation}
    \label{akt:emb}
    e_{(c_t, r_t)} = e_{c_t} + g_{r_t}
\end{equation}

Where $g_1$ and $g_0$ are embeddings for the correct and incorrect response, respectively. 


%

\section{\uppercase{Problem Statement}}
\label{problems}
We divide the problems addressed in this paper into two main parts: those that arise during evaluation and those that arise during training.

\subsection{Evaluation Problems}
\label{sec:eval}
\begin{figure}[h]
    \centering
\includegraphics[width=0.7\linewidth]{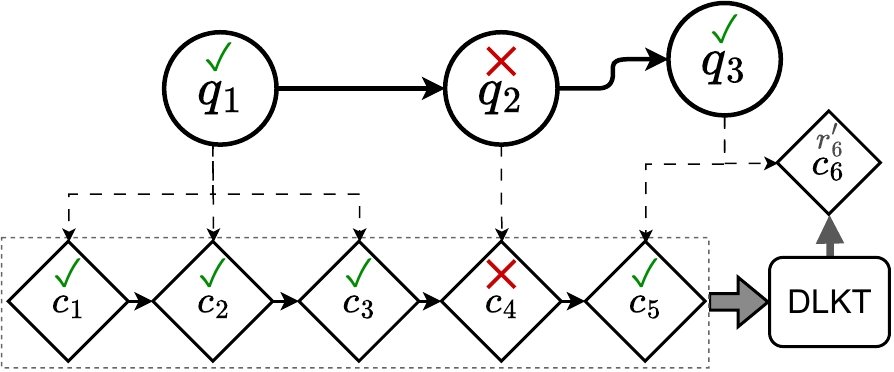}
\caption{\textit{one-by-one} evaluation and training on the expanded sequences. Note, $c_5$ ground-truth label can leak to $r_6'$ as both $c_5$ and $c_6$ belong to the same question, $q_3$. }\label{fig:onebyone} 
\end{figure}

In \cite{pykt}, the authors described two methods to evaluate models that operate on the expanded KC-student interaction sequence. They called the first method "one-by-one" evaluation, shown in Figure~\ref{fig:onebyone}, which involves evaluating the expanded sequence per KC, while ignoring the original question-student sequence.

\begin{figure}[h]
\includegraphics[width=0.7\linewidth]{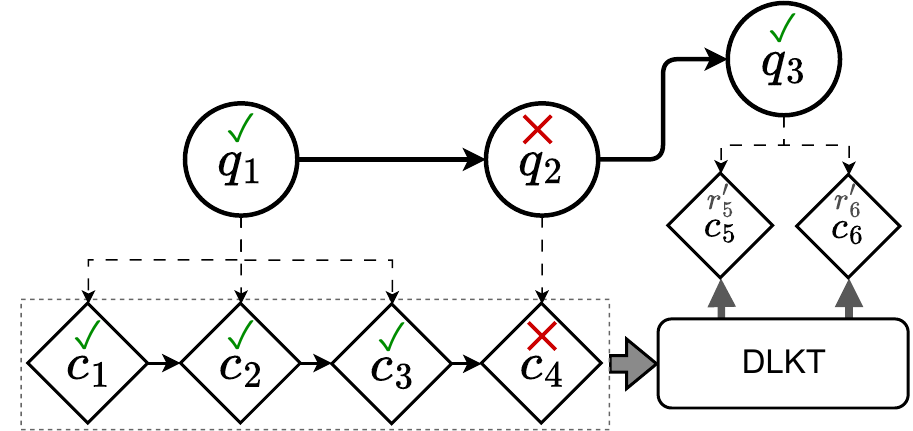}
\centering
\caption{The \textit{all-in-one} evaluation method. Both predictions of $c_5$ and $c_6$ should be produced independently of each other as they belong to the same question, $q_3$.}\label{allinone}
\end{figure}

The second method is what they call an "all-in-one" evaluation, which involves evaluating all KCs belonging to the same exercise at once, independently of each other, as shown in Figure~\ref{allinone}. Afterwards, outputs belonging to the same question are reduced using a chosen aggregation function, such as the mean, to represent the final prediction for the corresponding question. In this work, we always apply the mean as an aggregation method.

The "one-by-one" method does not match real production settings, as the ground-truth labels are not available for all KCs of the unanswered questions. This discrepancy produces misleading evaluation results \cite{pykt}. Thus, in this work, we enforce the "all-in-one" evaluation for all models that can leak ground-truth labels.

However, the "all-in-one" is expensive to compute since the sequence needs to be evaluated for each KC belonging to the same question independently before aggregating. Therefore, it is not practical to use the "all-in-one" method on the validation set, but only once on the test set. As these methods differ, validation can be misleading and may lead to the selection of an incorrect model when using methods such as cross-validation.


The second issue is that the expanded sequence is usually longer than the original sequence, and thus the available benchmarks should be used carefully. All models should be tested against the same number of questions per sequence for a fair comparison. However, since a maximum \textit{sequence window size} must be enforced, most implementations use the same window for both, expanded and original sequences, which leads to unfair comparisons since the expanded sequence contains fewer questions. This work enforces the requirement of consistent question window size across models to ensure fair comparison.   

%
%


\subsection {Training Problems}
\label{sec:train}

The "all-in-one" method can provide a reliable approach to evaluate the model to mimic production setting. However, during training, the expanded sequence still contains consecutive ground-truth labels for the same question. This can cause models to learn to leak ground-truth labels between KCs, leading to a deterioration in performance. This issue, along with the others discussed in the previous section, becomes more pronounced when dealing with datasets that contain a larger number of knowledge components per question, as will be demonstrated in section~\ref{resulsts}.

The problem stems from the way the expanded sequence is modeled, which does not match the distribution of the data during the production setting. Let $(c_t, r_t)$ and $(c_\tau, r_\tau)$ be two interactions in the expanded sequence, such that $\tau > t$ while $c_t$ and $c_\tau$ are two KCs. Let $B_{t,\tau}$ be the event that $t$ and $\tau$ belong to the same question in the sequence, $\{c_t, c_\tau\} \subset m(q_k)$. Let ${\sim}B_{t,\tau}$ be its complement. Let $H_\tau$ be all the interaction history before $\tau$, where $(c_t, r_t) \in H_\tau$, then the probability that $r_t$ and $r_\tau$ are equal can be modeled as follows

\begin{align}
P(r_\tau = r_t \mid H_\tau) &= P(r_\tau = r_t \mid  H_\tau, B_{t,\tau}) P(B_{t,\tau} \mid H_\tau) \nonumber\\ 
& + P(r_\tau = r_t \mid  H_\tau, {\sim}B_{t,\tau}) P({\sim}B_{t,\tau} \mid H_\tau) \nonumber\\ 
&\geq P(B_{t,\tau} \mid H_\tau) 
\end{align} 

The inequality holds because $P(r_\tau = r_t \mid H_\tau, B_{t,\tau})$ is one. A model trained on the expanded KC-student sequence can implicitly learn $P(B_{t,\tau} \mid H_\tau)$.  Consequently, with a high $P(B_{t,\tau} \mid H_\tau)$, the model can learn to ignore the history except for $r_t$. However, $B_{t,\tau}$ never occurs in production. This discrepancy between production and training will result in lower performance. Obviously, this is dataset dependent. For example, $B_{t,\tau}$ never occurs in a dataset with one KC per question.

Thus, our goal in this paper is to properly model the true distribution of the data by masking computation paths that can leak ground truth labels to accurately model production settings. We achieve that by ensuring that the sequence before the time $\tau$, $H_\tau$, does not contain $r_t$ if $B_{t,\tau}$ is true; instead, it should satisfy
\begin{equation}
H_\tau \subset \{(c_t, r_t) \mid t < \tau , {\sim}B_{t,\tau} \} \cup \{c_t \mid t < \tau, B_{t,\tau}\}
\end{equation}

which means we are dropping any ground-truth label that can leak to prediction at time $\tau$.

\section{\uppercase{Label Leakage-Free Framework}}
\label{main}

The goal is to remove the computation path between the responses to KCs of the same question in the expanded sequence. This idea shares some similarity with autoregressive models \cite{MADE}, where the computation path is masked to properly model the distribution using the autoregressive property of random variables to generate a true probability distribution. In our case, we mask computation paths that do not exist at production time and can cause ground-truth labels to leak.

If the model has no computation path between the responses that correspond to KCs of the same question then it does not suffer from label leakage during training or evaluation, which means they do not require an expensive "all-in-one" evaluation.

Note that it is possible to convert the one-to-many map, $m$, to a one-to-one mapping by treating each unique group of KCs that belong to the same question as a single KC. This approach is not feasible for models that rely on individual KC inputs, such as the Knowledge Query Network model \cite{kqn}. Additionally, for datasets with a high number of KCs per question, the resulting one-to-one mapping can have a large range. Finally, questions that have an unseen set of knowledge components are outside the scope of the one-to-one mapping. Therefore, we exclude this solution from our work.

\subsection{Incorporating a Mask Label}
\label{mlakt}

\begin{figure}[h]
    \centering
\includegraphics[width=0.7\linewidth]{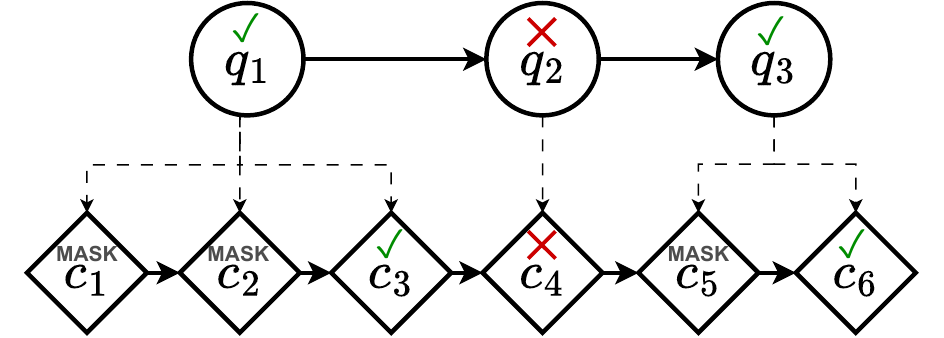}
\caption{Expanding question-student interaction sequence into KC-student interaction sequence with \texttt{MASK} labels. The green and red symbols are correct and incorrect respectively}\label{fig_mask}
\end{figure}

 We introduce a simple solution that can be applied to a wide variety of KT models. We simply add a mask label, $\texttt{MASK}$, to the expanded sequence alongside the correct $0$ and incorrect $1$ response labels. To prevent label leakage, we replace any ground truth label ($0$ and $1$) with \texttt{MASK} if it is followed by another KC of the same question in the sequence as shown in Figure \ref{fig_mask}. With this approach, each KC subsequence, that corresponds to a single question,  has only one ground-truth label at the end while the rest have \texttt{MASK} labels. 

  For example, let $q$ be a question in the sequence consisting of three KCs ($c_1, c_2, c_3$) and have a response, $r$, then it  would be represented in an expanded sequence as follows:

\[
\ldots, \left(q, c_1, \texttt{MASK}\right), \left(q, c_2, \texttt{MASK}\right), \left(q, c_3, r\right), \ldots
\]

Thus, the ground-truth $r$ can not be leaked, as it happens only at the last item in the sequence of KCs. Note that the \texttt{MASK} label is included only in the input sequence to the model, the output is still predicting only the original ground-truth labels, $0$ and $1$. This is applied during both training and inference.

Besides preventing label leakage, the mask label method adds explicit information about which KCs make up a question with a small increase in parameter size. This can be useful as other models, such as DKT, hide this information completely from the model.


Incorporating the new label into a model, usually does not need a huge change to the model itself, as we will see in the introduced model variants that use this method. To distinguish models that utilize this method from others, we append "-ML" to the model name, which stands for mask label. We introduce two model variants: mask label DKT (DKT-ML) and mask label AKT (AKT-ML).

\subsubsection{AKT-ML}

 As explained in section~\ref{akt}, AKT uses two separate vector embeddings for correct and incorrect labels, denoted as $g_1$ and $g_0$, respectively. Thus, to incorporate a \texttt{MASK} label, we only introduce a similar embedding $g_{\texttt{\scriptsize MASK}}$ with the same dimensions as $g_1$ and $g_0$. Its value is also learned during training similar to $g_1$ and $g_0$. 
 
 Building on that, a mask embedding for $(c_t, \texttt{MASK})$  is 

\begin{equation}
 e_{(c_t, \texttt{\scriptsize MASK})} = e_{c_t} + g_{\texttt{\scriptsize MASK}}
\end{equation}

 Furthermore, we need a new variation vector embedding (see \ref{akt}) for each KC-\texttt{MASK} pair, $(c_t, \texttt{MASK})$. Each unique pair can be mapped to a unique variation vector embedding, $f_{(c_t, \texttt{\scriptsize MASK})}$, to account for the new \texttt{MASK} label, which is done similarly to $f_{(c_t, 0)}$ and $f_{(c_t, 1)}$. With that, we can compute $e_{(\texttt{\scriptsize MASK}, c_t, q_t)}$ using equation \ref{eq:akt_emb2}.

\subsubsection{DKT-ML}
DKT discards questions from the sequence and thus only represents a separate embedding for each KC-response pair, $e_{(c_t, r_t)}$. One method to incorporate a \texttt{MASK} label is by adding a new embedding for each concept-\texttt{MASK} pair, $(c_t, \texttt{MASK})$.  However, we instead adapt an approach similar to AKT by using separate embeddings for each KC, denoted as $e_{c_t}$, and separate embeddings for each label: $g_0$, $g_1$, and $g_{\texttt{\scriptsize MASK}}$. Thus, the embedding for $(c_t, \texttt{MASK})$ is 

\begin{equation}
e_{(c_t, \texttt{\scriptsize MASK})} = e_{c_t} + g_{\texttt{\scriptsize MASK}}
\end{equation}

which can be passed to the RNN model.

\subsection{DKT with Averaged Embeddings}

Another approach to avoid label leakage is averaging all the embeddings of the constituting KCs before passing them to the model. This allows the model to avoid operating sequentially on the expanded sequence but instead on the original question-student interaction sequence, thus, avoiding all the mentioned problems. We created a DKT variant that uses this method, and we call it \textit{DKT-Fuse}.

DKT-Fuse adds a component that averages all the concept-response pair embeddings, $e_{(c, r)}$, of a specific question, $q$, before passing it to the RNN of DKT. A similar component has been used in the question-centric interpretable KT (QIKT) \cite{qikt} model. The output of the component can be described as follows:

\begin{equation}
\bar{e}_{(q, r)} = \frac{1}{|m(q)|} \sum_{c \in m(q)} e_{(c, r)}
\end{equation}

where  $m(q)$ is the set of KCs belonging to that question, while $\bar{e}_{(q, r)}$ is the average of the corresponding input embeddings. 

As we described in section~\ref{dkt}, DKT outputs a vector of probabilities corresponding to each KC in the dataset. Since we chose the mean as an aggregation method, the prediction of DKT-Fuse is the mean of the probabilities of the constituting KCs, $m(q)$, for each question $q$ during both training and evaluation.

\subsection{Special Attention Masks}

We can modify attention masks to cut any computation path between KCs belonging to the same question, and thus prevent label leakage. We apply these masks to AKT, and we refer to the model by question masked AKT (AKT-QM).
 
AKT-QM have the same architecture as AKT \cite{akt}. We only adjust the strictly lower triangular attention mask of AKT (described in section~\ref{akt}). To prevent peeking into KCs of the same question in the KC-question sequence. To do that, we replace the strictly lower triangular mask with 

\begin{equation}
A_{ij} = \left\{
\begin{array}{ll}
0 & \mbox{if } B_{i,j}\\
0 & \mbox{if } i <= j \\
1 & \mbox{otherwise}
\end{array}
\right.
\end{equation}

Where $B_{i,j}$ means $c_i$ and $c_j$ belong to the same question in the original sequence. Note that zero is mapped to $-\infty$ and one is mapped to zero when computing attention weights, see \cite{transformer}. These masks are computed using fancy indexing \cite{numpy,aaai2nd}. 

\subsection{Autoregressive Decoding of the DKT Model}

One method to avoid label leakage is to use autoregressive decoding for items belonging to the same question by sampling from the model instead of taking the ground-truth if a KC-response item can leak. This solution is computationally tractable, as ground-truth labels can be replaced with the model output at each recurrent step while parsing the sequence. We implement this for DKT and call it \textit{Autoregressive Decoding DKT} (DKT-AD).

DKT-AD employs the same architecture as DKT, with the distinction that the recurrent model substitutes ground-truth response values with model response predictions, $r'_t$, at time $t$ if $c_{t+1}$ and $c_t$ belong to the same question in the sequence. These samples are treated as constants, and thus no gradient propagation is performed.

To illustrate this approach, let $q$ be a question comprising three KCs, $c_1, c_2, c_3$, and eliciting a response $r$, then the model parses the following sequence:
$$
\ldots,\left(q, c_1, r'_1\right), \left(q, c_2, r'_2\right),(q, c_3, r),\ldots
$$
where $r$ is the ground-truth response data of the question, $q$, while $r'_1$ and $r'_2$ are model predictions. Each prediction depends on the preceding input sequence. 

\section{\uppercase{Experiments}}
\label{resulsts}
\begin{table*}[ht]
\centering
\caption{Dataset attributes after preprocessing. ques., KCs, studs., KCs/ques. stands for the number of questions, number of KCs, and average KCs per question respectively. KC-grps. stands for the number of unique KC groups assigned to each question.}\label{tab:datasets}
\begin{tabular}{llllll}
\toprule
\textbf{dataset} & ques. \phantom & KCs \phantom & studs. \phantom & KC-grps. \phantom  & KCs/ques. \\ \midrule
Algebra2005        & 173650          & 112           & 574               & 263                & 1.353                    \\
ASSISTments2009         & 17751           & 123           & 4163              & 149                & 1.196                    \\
CorrAS09    & 17751           & 246           & 4163              & 149                & 2.393                    \\
Duolingo2018       & 694675          & 2521          & 2638              & 7883               & 2.702                    \\ 
Riiid2020       & 13522          & 188          & 3822              & 1519               & 2.291           \\ 
\bottomrule
\end{tabular}
\end{table*}



As our focus is on models that utilize KCs, we chose datasets that have different KC related attributes such as KC cardinality, and the average KCs per question. We also list the number of unique KC groups in each dataset, where each group corresponds to the set of KCs belonging to a unique question. These attributes are shown in Table~\ref{tab:datasets}. Further, we discard all extra features and leave only KCs, questions, student identifiers and the order of interactions. We mainly use the following datasets:

\begin{itemize}
        \item \textbf{ASSISTments2009}\footnote{ Can be fetched from \url{https://sites.google.com/site/assistmentsdata/home/}}: Collected from the ASSISTments online tutoring platform between 2009 and 2010. We use the skill-builder version only.
        \item \textbf{Riiid2020} \cite{riiid}: Collected from an AI tutor. It contains more than 100 million student interaction. We only take the first one million interaction from this dataset. Moreover, it's worth noting that the Riiid2020 dataset was utilized in a competition context, permitting the incorporation of additional features, which we have deliberately omitted to ensure a fair comparison.
        \item \textbf{Algebra2005} \cite{algebra05}:  This data was part of the KDD Cup 2010 EDM Challenge. We only choose the data collected betweeen 2005-2006, titled "Algebra I 2005-2006".
        \item  \textbf{Duolingo2018}\cite{duolingo18}: Collected using Duolingo, an online language-learning app. It contains around $6k$ learners on different languages. We choose only students with English background learning Spanish. Further, we choose word tokens as KCs. Note that a word might consists of multiple tokens.
\end{itemize}


In order to emphasise the effect of label leakage during training, we further process the \textit{ASSISTments2009} dataset to contain perfectly correlated KCs. We achieve this by replacing each KC, $c$, with $m'(c)$ and $m''(c)$, where $m'$ and $m''$ are functions with disjoint ranges of cardinality equal to the original number of KCs. This means that each question has at least two KCs that are perfectly correlated (since they are duplicates). We abbreviate the generated dataset with \textit{CorrAS09}.

\subsection{Baselines and Training Setup}

To demonstrate the effect of preventing label leakage, we compare the proposed model variations in section~\ref{main} with their original counterparts: \textbf{AKT} and  \textbf{DKT}.

We also introduce baselines that do not expand KCs and therefore avoid the label-leakage problem. The following baselines are:

\begin{itemize}
    \item \textbf{DKVMN}\cite{dkvmn} is a memory augmented neural network model. The original implementation does not utilize KCs\cite{dkvmn,dkvmnca}
    \item \textbf{DeepIRT}\cite{deepirt}: Similar to DKVMN but it incorporate item response theory (IRT)\cite{irt} into the model to improve explainability. It also does not utilize KCs.
    \item \textbf{QIKT}\cite{qikt}: The model extracts knowledge from both questions and KCs before using IRT to output predictions to improve interpretability.
\end{itemize}

All models were trained with the ADAM optimizer \cite{adam} with learning rate $10^{-3}$. DKT based models were trained with a batch size of $128$ while all others were trained with a batch size of $24$. For each dataset, we perform 5-fold cross validation on $80\%$ of the students. We hold $20\%$ of the students as a test set. The metric used to assess performance is the area under the receiver operating characteristics curve (AUC)

All models were tested at the end of each fold on the same leave-out test set. Moreover, all models were tested at the question-level. For models that use an expanded sequence, the prediction for each question is calculated as the mean of the model’s outputs across all the KCs that constitute the question.

DKT and AKT are tested with the "all-in-one" method. However, they are chosen based on a "one-by-one" evaluation on the validation set during training which is what all implementations do according to our knowledge. This is because performing "all-in-one" evaluation is expensive to compute and using it for validation is not practical. This creates a divergence between validation and testing that can result in choosing the wrong model as explained in section~\ref{sec:eval}.

\begin{table*}[!ht]
\caption{AUC performance results across different model types. The marker \dag{} denotes models performing very close to the best-performing model, indicating a near tie. The marker * indicates not all folds have been tested (one fold is train 64\%, validation 16\%, test 20\%). The marker ** indicates impractical test on a dataset with a high number of questions for models that have question embeddings, they can be ignored.
}
\label{tab:fullbench}
\resizebox{\textwidth}{!}{%
\centering
\begin{tabular}{llllll}
\toprule
 & ASSISTments09 & CorrAS09 & Algebra05 & Riiid20 & Duolingo2018 \\
\midrule
\textbf{DKT} & $0.6990\pm0.0007$ & $0.6312\pm0.0014$ & $0.8070\pm0.0004$ & $0.5961\pm0.0003$ & $0.6518\pm0.0013$ \\
\textbf{DKT-ML} & $0.7185\pm0.0003$ & $0.7163\pm0.0006$ & $0.8178\pm0.0003$ & $0.6568\pm0.0004$ & $0.8681\pm0.0004$ \\
\textbf{DKT-AD} & $0.7180\pm0.0005$ & $0.7148\pm0.0002$ & $0.8161\pm0.0003$ & $0.6554\pm0.0003$ & $0.8679\pm0.0003$ \\
\textbf{DKT-Fuse} & $0.7066\pm0.0005$ & $0.7074\pm0.0008$ & $0.8175\pm0.0003$ & $0.6491\pm0.0003$ & $0.8786\pm0.0001$ \\
\textbf{AKT} & $0.7334\pm0.0017$ & $0.6361\pm0.0020$ & $0.7591\pm0.0045$ & $0.6136\pm0.0015$ & $0.7017\pm0.0183$ \\
\textbf{AKT-ML} & \bm{$0.7543\pm0.0010$} & \bm{$0.7552\pm0.0010$} & $\mathit{0.8282\pm0.0011}$\textsuperscript{\dag} & \bm{$0.7411\pm0.0007$} & \bm{$0.8807\pm0.0013$} \\
\textbf{AKT-QM} & $0.7193\pm0.0134$ & $0.7368\pm0.0011$ & $0.7919\pm0.0072$\textsuperscript{*} & $0.7289\pm0.0034$ & $0.8052$\textsuperscript{*} \\
\textbf{QIKT} & $0.7472\pm0.0008$ & $0.7484\pm0.0007$ & \bm{$0.8290\pm0.0007$} & $0.7306\pm0.0005$ & \multicolumn{1}{c}{$-$\textsuperscript{**}} \\
\textbf{DeepIRT} & $0.7215\pm0.0010$ &  $0.7215\pm0.0010$ & $0.7779\pm0.0003$ & $0.7312\pm0.0002$ & $0.5294\pm0.0002$\textsuperscript{**} \\
\textbf{DKVMN} & $0.7215\pm0.0011$ & $0.7215\pm0.0011$  & $0.7768\pm0.0003$ & $0.7327\pm0.0003$ & $0.5296\pm0.0001$\textsuperscript{**}\\
\bottomrule
\end{tabular}}
\end{table*}

On the other hand, all introduced model variants in section~\ref{main} do not need an "all-in-one" evaluation. They use the same method for both validation and testing, which is to use the output of the basic "one-by-one" evaluation but aggregating the results per question by finding the mean of the constituting KCs.

\subsection{Results and Discussion}

In order to shed light on the problem of label leakage during training (section~\ref{sec:train}), we compared the original implementations of AKT and DKT to the model variants introduced in section~\ref{main} using the ASSISTments2009 and CorrAS09 datasets. CorrAS09 is identical to ASSISTments2009, but with duplicate KCs. Table~\ref{tab:fullbench} shows that both AKT and DKT experience a significant decrease in performance on CorrAS09, unlike all the introduced model variants. 

\begin{figure*}
    \centering
\includegraphics[width=0.95\linewidth]{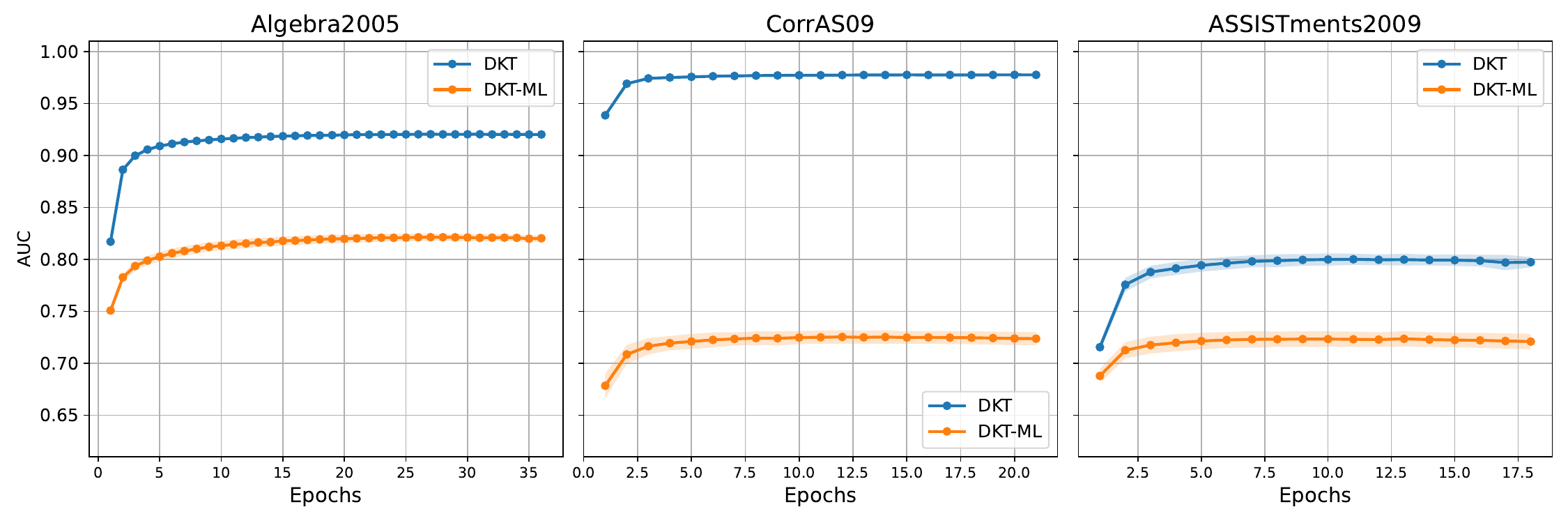}\hfill
\caption{Validation loss on the CorrAS09, Algebra2005, and ASSISTments2009 datasets. The "one-by-one" method is used for the original DKT model and DKT-ML, our variant of DKT with added mask label. DKT has inflated results as it simply learned to leak labels. This becomes more pronounced for datasets with higher KCs per question (Algebra2005 and CorrAS09). DKT-ML demonstrates similar performance on CorrAS09 and ASSISTments2009.}\label{fig:train}
\end{figure*}

Furthermore, the introduced models greatly outperform their original counterparts on the Duolingo2018 and Riiid2020 datasets as seen in Table~\ref{tab:fullbench}. Both of these datasets have high average number of KCs per question. Moreover, Riiid2020 has the highest count of unique KC groups relative to its question count. This suggest a strong negative effect of label leakage, which becomes more pronounced with an increase in the number of KCs per question. The introduced models are also outperforming their original counterpart on datasets with few KCs per question. However, the label leakage effect is not as significant.

Finally, models trained with a mask label, \texttt{MASK}, exhibit competitive performance. Specifically, AKT-ML is winning in nearly all benchmarks, even surpassing models that do not expand knowledge concepts (KCs).

All benchmark results were tested with a window size of $150$ questions, which is a fixed window size unless the sequence does not contain $150$ questions. If the model requires an expanded KC-student sequence, the $150$ questions are further processed into an expanded sequence of KC-student interactions, thus avoiding inconsistent sequence lengths between models that expand the original sequences into KCs and models that do not (\textbf{KDVMN}, \textbf{QIKT}, \textbf{DeepIRT}). Note that the $150$ questions can be expanded to around $400$ in a dataset like Duolingo2018. If we do not enforce this, we can falsely increase performance results. For example, DKT has a mean AUC of $0.6622$ over a fixed expanded sequence length of $150$ KCs instead of $150$ question on the Duolingo2018 dataset, which is an improvement over the reported value of $0.6518$ in Table~\ref{tab:fullbench}. Similarly, it has a mean AUC of $0.815$ for the Algebra2005 dataset, which is higher than the reported value of $0.8070$.



Lastly, as we mentioned in section \ref{problems}, the expensive to compute "all-in-one" method that mimics a true production environment for these models diverges from the validation loss used to choose these models during training ("one-by-one" method). These validation results are misleading due to the label leakage problem. For example, the AUC validation results during training on the CorrAS09 are extremely higher than the ASSISTments2009 dataset despite having the same question data, see Figure \ref{fig:train}. 

To encourage reproducibility, the results of this work can be reproduced using our open-source tool, available at \url{https://github.com/badranx/KTbench}.





\section{\uppercase{Conclusions}}
In this work, we identified a ground-truth label leakage problem in KT models that are trained on the expanded KC-student interaction sequence. Given the importance of these models, we introduced a number of methods to avoid the label leakage problem. Our model variants that use these methods outperformed their original counterparts, and some showed competitive performance in general.

The importance of this work is to shed light on the problem of label leakage and to influence future KT model architectures to avoid certain design choices that can lead to this problem,  especially when dealing with data that has a relatively high average number of KCs per item.

\section*{\uppercase{Acknowledgements}}
We would like to thank the anonymous reviewers for their valuable comments and constructive feedback. This work was funded by the federal state of Baden-Württemberg as part of the Doctoral Certificate Programme "Wissensmedien" (grant number BW6{\_}10).
\newline

\bibliographystyle{apalike}
{\small
\bibliography{example}}

\begin{thebibliography}{}

\bibitem[Abdelrahman et~al., 2023]{abdelrahman2023knowledge}
Abdelrahman, G., Wang, Q., and Nunes, B. (2023).
\newblock Knowledge tracing: A survey.
\newblock {\em ACM Comput. Surv.}, 55(11).

\bibitem[Ai et~al., 2019]{dkvmnca}
Ai, F., Chen, Y., Guo, Y., Zhao, Y., Wang, Z., Fu, G., and Wang, G. (2019).
\newblock Concept-aware deep knowledge tracing and exercise recommendation in
  an online learning system.
\newblock In {\em Proceedings of the 12th International Conference on
  Educational Data Mining}, pages 240--245, Montr{\'{e}}al, Canada.
  International Educational Data Mining Society {(IEDMS)}.

\bibitem[Chen et~al., 2023]{qikt}
Chen, J., Liu, Z., Huang, S., Liu, Q., and Luo, W. (2023).
\newblock Improving interpretability of deep sequential knowledge tracing
  models with question-centric cognitive representations.
\newblock {\em Proceedings of the AAAI Conference on Artificial Intelligence},
  37(12):14196--14204.

\bibitem[Choi et~al., 2020]{riiid}
Choi, Y., Lee, Y., Shin, D., Cho, J., Park, S., Lee, S., Baek, J., Bae, C.,
  Kim, B., and Heo, J. (2020).
\newblock Ednet: A large-scale hierarchical dataset in education.
\newblock In {\em International Conference on Artificial Intelligence in
  Education}, pages 69--73, Morocco. Springer.

\bibitem[Corbett and Anderson, 1994]{bkt}
Corbett, A.~T. and Anderson, J.~R. (1994).
\newblock Knowledge tracing: Modeling the acquisition of procedural knowledge.
\newblock {\em User modeling and user-adapted interaction}, 4:253--278.

\bibitem[Devlin et~al., 2019]{BERT}
Devlin, J., Chang, M., Lee, K., and Toutanova, K. (2019).
\newblock {BERT:} pre-training of deep bidirectional transformers for language
  understanding.
\newblock In Burstein, J., Doran, C., and Solorio, T., editors, {\em
  Proceedings of the 2019 Conference of the North American Chapter of the
  Association for Computational Linguistics: Human Language Technologies,
  {NAACL-HLT} 2019, Minneapolis, MN, USA, June 2-7, 2019, Volume 1 (Long and
  Short Papers)}, pages 4171--4186. Association for Computational Linguistics.

\bibitem[Germain et~al., 2015]{MADE}
Germain, M., Gregor, K., Murray, I., and Larochelle, H. (2015).
\newblock Made: Masked autoencoder for distribution estimation.
\newblock In {\em Proceedings of the 32nd International Conference on Machine
  Learning}, volume~37 of {\em Proceedings of Machine Learning Research}, pages
  881--889, Lille, France. PMLR.

\bibitem[Gervet et~al., 2020]{whendeep}
Gervet, T., Koedinger, K., Schneider, J., Mitchell, T., et~al. (2020).
\newblock When is deep learning the best approach to knowledge tracing?
\newblock {\em Journal of Educational Data Mining}, 12(3):31--54.

\bibitem[Ghosh et~al., 2020]{akt}
Ghosh, A., Heffernan, N., and Lan, A.~S. (2020).
\newblock Context-aware attentive knowledge tracing.
\newblock In {\em Proceedings of the 26th ACM SIGKDD International Conference
  on Knowledge Discovery \& Data Mining}, KDD '20, page 2330–2339, New York,
  NY, USA. Association for Computing Machinery.

\bibitem[Harris et~al., 2020]{numpy}
Harris, C.~R., Millman, K.~J., van~der Walt, S., Gommers, R., Virtanen, P.,
  Cournapeau, D., Wieser, E., Taylor, J., Berg, S., Smith, N.~J., Kern, R.,
  Picus, M., Hoyer, S., van Kerkwijk, M.~H., Brett, M., Haldane, A., del
  R{\'{\i}}o, J.~F., Wiebe, M., Peterson, P., G{\'{e}}rard{-}Marchant, P.,
  Sheppard, K., Reddy, T., Weckesser, W., Abbasi, H., Gohlke, C., and Oliphant,
  T.~E. (2020).
\newblock Array programming with numpy.
\newblock {\em Nat.}, 585:357--362.

\bibitem[Khajah et~al., 2016]{Howdeep}
Khajah, M.~M., Lindsey, R.~V., and Mozer, M.~C. (2016).
\newblock How deep is knowledge tracing?
\newblock {\em ArXiv}, abs/1604.02416.

\bibitem[Kingma and Ba, 2015]{adam}
Kingma, D. and Ba, J. (2015).
\newblock Adam: A method for stochastic optimization.
\newblock In {\em 3rd International Conference on Learning Representations},
  San Diega, CA, USA.

\bibitem[Lee and Yeung, 2019]{kqn}
Lee, J. and Yeung, D.-Y. (2019).
\newblock Knowledge query network for knowledge tracing: How knowledge
  interacts with skills.
\newblock In {\em Proceedings of the 9th International Conference on Learning
  Analytics \& Knowledge}, LAK19, page 491–500, New York, NY, USA.
  Association for Computing Machinery.

\bibitem[Liu et~al., 2022]{pykt}
Liu, Z., Liu, Q., Chen, J., Huang, S., Tang, J., and Luo, W. (2022).
\newblock pykt: A python library to benchmark deep learning based knowledge
  tracing models.
\newblock In {\em Advances in Neural Information Processing Systems},
  volume~35, pages 18542--18555. Curran Associates, Inc.

\bibitem[Mao, 2018]{BKTindep}
Mao, Y. (2018).
\newblock Deep learning vs. bayesian knowledge tracing: Student models for
  interventions.
\newblock {\em Journal of educational data mining}, 10(2).

\bibitem[Nagatani et~al., 2019]{DKTforget}
Nagatani, K., Zhang, Q., Sato, M., Chen, Y., Chen, F., and Ohkuma, T. (2019).
\newblock Augmenting knowledge tracing by considering forgetting behavior.
\newblock In Liu, L., White, R.~W., Mantrach, A., Silvestri, F., McAuley,
  J.~J., Baeza{-}Yates, R., and Zia, L., editors, {\em The World Wide Web
  Conference, {WWW} 2019, San Francisco, CA, USA, May 13-17, 2019}, pages
  3101--3107. {ACM}.

\bibitem[Oya and Morishima, 2021]{aaai2nd}
Oya, T. and Morishima, S. (2021).
\newblock Lstm-sakt: Lstm-encoded sakt-like transformer for knowledge tracing.

\bibitem[Piech et~al., 2015]{dkt}
Piech, C., Bassen, J., Huang, J., Ganguli, S., Sahami, M., Guibas, L.~J., and
  Sohl-Dickstein, J. (2015).
\newblock Deep knowledge tracing.
\newblock {\em Advances in neural information processing systems}, 28.

\bibitem[Rasch, 1993]{irt}
Rasch, G. (1993).
\newblock {\em Probabilistic models for some intelligence and attainment
  tests.}
\newblock ERIC.

\bibitem[Settles et~al., 2018]{duolingo18}
Settles, B., Brust, C., Gustafson, E., Hagiwara, M., and Madnani, N. (2018).
\newblock Second language acquisition modeling.
\newblock In {\em Proceedings of the thirteenth workshop on innovative use of
  NLP for building educational applications}, pages 56--65, New Orleans, LA,
  USA. Association for Computational Linguistics.

\bibitem[Stamper et~al., 2010]{algebra05}
Stamper, J., Niculescu-Mizil, A., Ritter, S., Gordon, G., and Koedinger, K.
  (2010).
\newblock [data set name]. [challenge/development] data set from kdd cup 2010
  educational data mining challenge.
\newblock Retrieved from
  \url{http://pslcdatashop.web.cmu.edu/KDDCup/downloads.jsp}.

\bibitem[Vaswani et~al., 2017]{transformer}
Vaswani, A., Shazeer, N., Parmar, N., Uszkoreit, J., Jones, L., Gomez, A.~N.,
  Kaiser, L.~u., and Polosukhin, I. (2017).
\newblock Attention is all you need.
\newblock In {\em Advances in Neural Information Processing Systems},
  volume~30. Curran Associates, Inc.

\bibitem[Wu et~al., 2025]{edustudio}
Wu, L., Chen, X., Liu, F., Xie, J., Xia, C., Tan, Z., Tian, M., Li, J., Zhang,
  K., Lian, D., et~al. (2025).
\newblock Edustudio: towards a unified library for student cognitive modeling.
\newblock {\em Frontiers of Computer Science}, 19(8):198342.

\bibitem[Yeung, 2019]{deepirt}
Yeung, C. (2019).
\newblock Deep-irt: Make deep learning based knowledge tracing explainable
  using item response theory.
\newblock In {\em Proceedings of the 12th International Conference on
  Educational Data Mining}, Montr{\'{e}}al, Canada. International Educational
  Data Mining Society {(IEDMS)}.

\bibitem[Yeung and Yeung, 2018]{dkt_plus}
Yeung, C. and Yeung, D. (2018).
\newblock Addressing two problems in deep knowledge tracing via
  prediction-consistent regularization.
\newblock In {\em Proceedings of the Fifth Annual {ACM} Conference on Learning
  at Scale}, pages 5:1--5:10, London, UK. {ACM}.

\bibitem[Zhang et~al., 2017]{dkvmn}
Zhang, J., Shi, X., King, I., and Yeung, D. (2017).
\newblock Dynamic key-value memory networks for knowledge tracing.
\newblock In {\em Proceedings of the 26th International Conference on World
  Wide Web}, pages 765--774, Perth, Australia. {ACM}.

\end{thebibliography}

\end{document}